\documentclass[prd,showpacs,preprintnumbers,amsmath,amssymb]{revtex4}
\usepackage{graphicx}
\usepackage{dcolumn}
\usepackage{bm}
\usepackage{epsfig}

\def\calm{{\cal M}}

\def\ran{\rangle}

\def\aq{\bar{q}}
\def\beq{\begin{equation}}
\def\eeq{\end{equation}}

\def\bfig{\begin{figure}}
\def\efig{\end{figure}}
\def\bea{\begin{eqnarray}}
\def\eea{\end{eqnarray}}
\def\bwt{\begin{widetext}}
\def\ewt{\end{widetext}}
\def\beann{\begin{eqnarray*}}
\def\eeann{\end{eqnarray*}}

\def\nn{\nonumber}

\def\3p0{$^{3}P_{0}$}

\begin{document}

\title{LIGHT MESON DECAY IN THE  C$^{3}P_{0}$ MODEL}

\author{J. N. DE QUADROS, D. T.  DA SILVA, M. L. L. DA SILVA
and D. HADJIMICHEF}
\email{dimihadj@gmail.com, dimiter.hadjimichef@ufrgs.br}
\affiliation{Instituto de F\'{\i}sica, Universidade Federal do Rio
  Grande do  Sul\\
 Av. Bento Gon\c{c}alves, 9500, Porto Alegre, Rio Grande do Sul,
CEP 91501-970, Brazil}


\begin{abstract}

Having its origin in a successful mapping technique,
the Fock-Tani formalism, the corrected  $^{3}P_{0}$ model (C$^{3}P_{0}$) retains the basic aspects of the \3p0 predictions with the
inclusion of bound-state corrections. Evaluation of the decay amplitudes has be performed for open-flavor strong decays
in the light meson sector. The bound-state corrections introduce a fine-tuning for the former \3p0 model, in particular,
the adjustment of the $D/S$  ratios in $b_{1}\rightarrow\omega\pi$, $a_{1}\rightarrow\rho\pi$ and
$h_1\to\rho\pi$ decays.


\end{abstract}


\maketitle


\section{Introduction}
\label{intro}

A  mapping technique
long used in atomic physics \cite{girar1}-\cite{girar3}, the Fock-Tani formalism
(FTf), has been adapted, in previous publications \cite{annals}-\cite{mario}, in order
to describe hadron-hadron scattering interactions with constituent
interchange. Recently   this technique has been extended  to
 meson decay \cite{prd08}. The novel feature of this approach is
 the presence of bound-state  corrections (BSC) in the decay amplitude. A necessary
ingredient in the formalism is the definition of the microscopic interaction Hamiltonian
between the elementary constituents.  

Open-flavor strong decays are successfully described in the context of the 
 the \3p0 model, which considers only OZI-allowed strong-interaction decays and was introduced over thirty years ago by Micu
\cite{micu}-\cite{leyaouanc5}. In the FTf, if one  starts from a microscopic $q\bar{q}$ pair-creation interaction,
in lowest order, the \3p0 results are reproduced. In higher orders of the formalism corrections due to
the bound-state nature of the mesons are present and the
 $q\bar{q}$ interaction strength is modified. This new model is called the {\it Corrected \3p0 model} (C\3p0) \cite{prd08}.
 Light meson decay and other meson sectors have been 
studied by  T. Barnes {\it  et al.} \cite{barnes1}-\cite{barnes4}
with the \3p0 model. In their formulation two basic parameters are adjusted to data,
$\gamma$ (the interaction strength)  and $\beta$ (the wave function's
extension parameter). They found  optimum values for these parameters near
$\gamma= 0.5$ and $\beta=0.4$ GeV.

In the present work, we employ the FTf to the light 1S and 1P  decays and a comparison
is made with the usual \3p0 results.
In the next section we briefly review the basic aspects of the formalism and the
C\3p0 derivation.
Section III is dedicated to obtain the decay rates of seven light mesons, followed
by the summary and conclusions.


\section{The C$^{3}P_{0}$ model}
\label{sec:mesons}

In this section we present a brief review of the formal aspects regarding the Fock-Tani mapping procedure and
how it is implemented  to quark-antiquark meson states \cite{annals,plb}.
The starting point of the Fock-Tani formalism  is the definition of single
composite bound states. We write a single-meson state in terms of a meson
creation operator $M_{\alpha}^{\dagger}$ as
\bea
|\alpha \ran  = M_{\alpha}^{\dagger}|0 \ran ,
\label{1b}
\eea
where $|0 \ran$ is the vacuum state. The meson creation operator $M_{\alpha}^{\dagger}$ is written
in terms of constituent quark and antiquark creation operators
$q^{\dagger}$ and $\aq^{\dagger}$,
\bea
M^{\dagger}_{\alpha}= \Phi_{\alpha}^{\mu \nu}
q_{\mu}^{\dagger} {\aq}_{\nu}^{\dagger} ,
\label{Mop}
\eea
$\Phi_{\alpha}^{\mu \nu}$ is the meson wave function and $q_{\mu}|0
\ran=\aq_{\nu}|0\ran=0$. The index $\alpha$ identifies the meson
quantum numbers of space, spin and isospin. The indices $\mu$ and
$\nu$ denote the spatial, spin, flavor, and color quantum numbers of
the constituent quarks. A sum over repeated indices is implied. The
meson operators satisfy the following non-canonical commutation
relations
\beq
[M_{\alpha}, M^{\dagger}_{\beta}]=\delta_{\alpha \beta} -
M_{\alpha \beta},\hspace{1.5cm}[M_{\alpha}, M_{\beta}]=0,
\label{Mcom}
\eeq
where
\bea
M_{\alpha \beta}= \Phi_{\alpha}^{*{\mu \nu }}
\Phi_{\beta}^{\mu \sigma }\aq^{\dagger}_{\sigma}\aq_{\nu}
+ \Phi_{\alpha}^{*{\mu \nu }}
\Phi_{\beta}^{\rho \nu}q^{\dagger}_{\rho}q_{\mu}.
\label{delta}
\eea
A transformation is defined such that  a single-meson state
$|\alpha \ran$ is redescribed by an (``ideal") elementary-meson state by
\bea
|\alpha \ran\longrightarrow U^{-1}|\alpha\rangle =
m^{\dagger}_{\alpha}|0\rangle,
\label{single_mes}
\eea
where $m^{\dagger}_{\alpha}$ an ideal meson creation operator. The ideal
meson operators $m^{\dagger}_{\alpha}$ and $m_{\alpha}$ satisfy,
by definition, canonical commutation relations
\beq
[m_{\alpha}, m^{\dagger}_{\beta}]=\delta_{\alpha \beta} ,
\hspace{1.5cm}[m_{\alpha}, m_{\beta}]=0.
\label{mcom}
\eeq
Once a microscopic interaction Hamiltonian $H_I$ is defined, at the quark
level, a new transformed Hamiltonian can be obtained. This effective
interaction,  the {\sl Fock-Tani Hamiltonian} ($H_{\rm FT}$),  is
obtained   by the application of the unitary operator
$U$ on the microscopic Hamiltonian $H_I$, {\it i.e.}, $H_{\rm FT}=U^{-1}\,H_I\,U$.
The transformed Hamiltonian   describes all possible
processes involving mesons and quarks.
In $H_{\rm FT}$ there are higher order terms that provide bound-state corrections
to the lower order ones. The basic quantity for these corrections is the {\it bound-state kernel}
$\Delta$ defined as
\begin{eqnarray}
\Delta(\rho\tau;\lambda\nu)
=\Phi^{\rho\tau}_{\alpha}\Phi^{\ast\lambda\nu}_{\alpha}.
\label{kernel}
\end{eqnarray}
The physical meaning of the $\Delta$ kernel becomes evident, in the
sense that it   modifies  the quark-antiquark interaction strength \cite{annals,plb,prd08}.

In the present calculation, the microscopic interaction Hamiltonian
is a  pair creation Hamiltonian $H_{q\aq}$  defined as 
\bea H_{q\aq}=V_{\mu\nu}\,
q^{\dag}_{\mu}\aq^{\dag}_{\nu} \,,
\label{h_3p0} 
\eea 
where in (\ref{h_3p0})
 a sum (integration) is again  implied  over repeated indexes \cite{prd08}. 
%
The pair creation potential $V_{\mu\nu}$ is given by 
\bea 
V_{\mu\nu}\equiv
2\,m_{q}\, \gamma\, \,\delta(\vec{p}_{\mu}+\vec{p}_{\nu})\,
\bar{u}_{s_{\mu}f_{\mu}c_{\mu}  } (\vec{p}_{\mu}) \, v_{s_{\nu}
f_{\nu}c_{\nu}  }(\vec{p}_{\nu}) , 
\label{vmn} 
\eea 
where $\gamma$ is the pair production strength.
The pair production is obtained from the  non-relativistic limit 
 of $H_{q\aq}$ involving Dirac quark fields \cite{barnes1}.
Applying the Fock-Tani transformation to $H_{q\aq}$ one obtains the effective
Hamiltonian that describes a decay process.
In the FTf perspective a new
 aspect is introduced  to meson decay: bound-state corrections.
The lowest order correction is one that involves only one
bound-state kernel $\Delta$. 
The bound-state corrected, C\3p0 Hamiltonian, is
\bea
\!\!\!\!\!\!\!\!\!\!\!\!\!\!\!\!
H^{\rm C3P0}
&=& -\Phi^{\ast\rho\xi}_{\alpha} \Phi^{\ast\lambda\tau}_{\beta}
\Phi^{\omega\sigma}_{\gamma}\, V^{\rm C3P0}\, m^{\dag}_{\alpha}
m^{\dag}_{\beta} m_{\gamma},
\label{c3p0}
\eea
where $V^{\rm C3P0} $
is a condensed notation for \bea
 V^{\rm C3P0}&=&
\left[\delta_{\mu\lambda}
\delta_{\nu\xi}
\delta_{\omega\rho}
\delta_{\sigma\tau}
-
\frac{1}{2}
\delta_{\sigma\xi}\,\delta_{\lambda\omega}\,\,
\Delta(\rho\tau;\mu\nu)
+
\frac{1}{4}
\delta_{\sigma\xi}\,\delta_{\lambda\mu}\,\,
\Delta(\rho\tau;\omega\nu)
\right.
\nn\\
&&
+
\left.
\frac{1}{4}
\delta_{\xi\nu}\,\delta_{\lambda\omega}\,\,
\Delta(\rho\tau;\mu\sigma)
\right]V_{\mu\nu}.
\label{vc3p0}
\eea
In the ideal meson space the initial and final states involve only ideal
meson operators $|A\rangle=m^{\dag}_{\gamma}|0\rangle$ and 
$|BC\rangle=m^{\dag}_{\alpha}m^{\dag}_{\beta} |0\rangle $.
The C\3p0 amplitude is obtained by the following matrix element,
\bea 
\hspace{-1cm}
\langle B C | H^{\rm C3P0} | A \rangle &=&
\delta(\vec{P}_A-\vec{P}_B-\vec{P}_C)\, h_{fi}^{\rm C3P0}
\label{ideal-matrix} 
\eea
The $h_{fi}^{\rm C3P0}$ decay amplitude is combined with
relativistic phase space, resulting in the differential decay rate
\bea
\frac{d\Gamma_{A\to BC}}{d\Omega}=2\pi\,P\,
\frac{E_B\,E_C}{M_A}|h_{fi}^{\rm C3P0}|^2
\label{dif-gamma}
\eea
which, after integration in the solid angle $\Omega$, a  usual choice for the meson
momenta is made: $\vec{P}_A=0$
($P=|\vec{P}_B|=|\vec{P}_C|$).



\section{Light Meson Decay in the C\3p0 Model}

We shall apply the model described in the former section to 
the light meson sector,  in particular  1S and 1P decay processes:
$\rho\to\pi\pi$, $b_1\to\omega\pi$, $a_1\to\rho\pi$,
$a_2\to\rho\pi$, $h_1\to\rho\pi$, $f_0\to\pi\pi$ and
$f_2\to\pi\pi$. 
The major issue in this calculation is the value of the decay amplitude  $h_{fi}^{\rm C3P0}$.
To evaluate this quantity we must define  the general non-relativistic
 meson wave function $\Phi_{\alpha}^{\mu\nu}$,   which can be written as a direct
product 
\bea
\Phi_{\alpha}^{\mu\nu}=
\chi_{S_\alpha }^{ s_{1}s_{2} }
f_{f_{\alpha} }^{f_{1}f_{2}}
C^{c_{1}c_{2}}
\Phi_{nl }( \vec{P}_{\alpha}-\vec{p}_{1}-\vec{p}_{2})
\;,
\label{funcdomeson}
\eea
where the components are: spin  $\chi_{S_{\alpha}}^{s_{1}s_{2}}$
[ $s_{1}$ and $s_{2}$ are
the quark (antiquark) spin projections,
$S_{\alpha} $ denotes the meson spin];
flavor $f_{f_{\alpha}}^{f_{1}f_{2}}$; color $C^{c_{1}c_{2}}$ and
space $\Phi_{nl }(\vec{P}_{\alpha}-\vec{p}_{1}-\vec{p}_{2})$.
In all our calculations the color component will be given by
\bea
C^{c_{1}c_{2}}=\frac{1}{\sqrt{3}}\,\,\delta^{c_{1}c_{2}}.
\label{color}
\eea
We assume that the spatial part is defined as harmonic oscillator wave functions
\bea
\Phi_{nl }(\vec{P}_{\alpha}-\vec{p}_{1}-\vec{p}_{2})
=
\delta(\vec{P}_{\alpha}-\vec{p}_{1}-\vec{p}_{2})
\,\,\phi_{nl}(\vec{p}_1,\vec{p}_2),
\eea
where
$\phi_{nl}(\vec{p}_i,\vec{p}_j)  $ is given by
\bea
\phi_{nl}(\vec{p}_i,\vec{p}_j)&=&
 (\frac{1}{2\beta})^{l}\, N_{nl}\, |\vec{p}_i-\vec{p}_j|^{l}\,
\exp\left[ -\frac{(\vec{p}_i-\vec{p}_j)^2   }{ 8\beta^2 }   \right]\,
{\cal L}_{n}^{l+\frac{1}{2}}
\left[\frac{(\vec{p}_i-\vec{p}_j)^2   }{ 4\beta^2} \right]
 Y_{lm}(\Omega_{\vec{p}_i-\vec{p}_j}     ),
\label{psi_oh}
\eea
with $p_{i(j)} $ the internal momentum, the spherical harmonic $Y_{lm}$ and
$\beta$ a scale parameter. The normalization constant $N_{nl}$  dependent
on the radial and orbital quantum numbers
\bea
N_{nl}=
\left[
\frac{2 (n!)}{\beta^3\,\Gamma(n+l+3/2) }
\right]^{\frac{1}{2}}.
\eea
The Laguerre polynomials ${\cal L}_{n}^{l+\frac{1}{2}}(p)$ are defined as
\bea
{\cal L}_{n}^{l+\frac{1}{2}}(p)=\sum_{k=0}^{n}
\frac{(-)^k\, \Gamma(n+l+3/2)  }{k!\,(n-k)! \, \Gamma(k+l+3/2)  }\,\,p^k\,.
\label{laguerre}
\eea

The bound-state kernel's definition in (\ref{kernel})
implies in an additional element, due to the contraction in the
$\alpha$ index, a {\it sum over species}
requirement \cite{girar1}-\cite{girar3,prd08}.  A question that naturally arises is:  which
states to include in this sum?  We shall adopt in our
calculation a restrictive choice: include in the sum only
the particles that are present in the final state.
For the example, in the  $\rho^{+}\to\pi^{0}+\pi^{+}$, decay, $\Delta\left(\rho\tau;\lambda\nu\right)$
will have two contributions set to the quantum numbers of  $\pi^{0}$ and $\pi^{+}$.
In the $b_1^{+}\to\omega+\pi^{+}$ decay,  two contributions come from  $\omega$ and $\pi^{+}$. Similarly,
the $a_1^{+}\to\rho^{+}+\pi^{0}$ decay shall be corrected by the final state mesons $\rho^{+}$ and $\pi^{0}$. The
other decays follow the same logic. 

After calculating the matrix element in Eq. (\ref{ideal-matrix} ), the  general
decay amplitude can be written as 
\bea 
h_{fi}^{\rm C3P0} & = &
\frac{\gamma}{\pi^{1/4}\,\beta^{1/2}}
\,\calm_{fi}. 
\label{hfi-decay} 
\eea
where
\bea
\calm_{fi}^{\rho\to\pi\pi} &=&
{\cal C}^{\rho}_{10}\,
Y_{11}\left(\Omega_x\right)
\,\,\,\,\,\,\,\,\,\,\,\,\,\,\,\,\,\,\,\,\,\,\,\,\,\,\,\,\,
\nonumber \\
\calm_{fi}^{f2\to\pi\pi} &=&
{\cal C}^{f_2}_{20}\, Y_{22}\left(\Omega_x\right)
\,\,\,\,\,\,\,\,\,\,\,\,\,\,\,\,\,\,\,\,\,\,\,\,\,\,\,\,\,
\nonumber \\
\calm_{fi}^{f_0\to\pi\pi} &=&
{\cal C}^{f_0}_{00}\, Y_{00}\left(\Omega_x\right)
\,\,\,\,\,\,\,\,\,\,\,\,\,\,\,\,\,\,\,\,\,\,\,\,\,\,\,\,\,
\nonumber \\
\calm_{fi}^{a_2\to\rho\pi} &=&
{\cal C}^{a_2}_{21}\, Y_{21}\left(\Omega_x\right)
\,\,\,\,\,\,\,\,\,\,\,\,\,\,\,\,\,\,\,\,\,\,\,\,\,\,\,\,\,
\nonumber \\
\calm_{fi}^{b1\to\omega\pi}  &=&
{\cal C}^{b_1}_{01}\, Y_{00} \left(\Omega_x\right)
+{\cal C}^{b_1}_{21}\, Y_{20}\left(\Omega_x\right)
\,\,\,\,\,\,\,\,\,\,\,\,\,\,\,\,\,\,\,\,\,\,\,\,\,\,\,\,\,
\nonumber \\
\calm_{fi}^{a_1\to\rho\pi}  &=&
{\cal C}^{a_1}_{01}\, Y_{00} \left(\Omega_x\right)
+{\cal C}^{a_1}_{21}\, Y_{20}\left(\Omega_x\right)
\,\,\,\,\,\,\,\,\,\,\,\,\,\,\,\,\,\,\,\,\,\,\,\,\,\,\,\,\,
\nonumber \\
\calm_{fi}^{h_1\to\rho\pi} &=&
{\cal C}^{h_1}_{01}\, Y_{00} \left(\Omega_x\right)
+{\cal C}^{h_1}_{21}\, Y_{20}\left(\Omega_x\right),
\label{hfic3po-rho7}
\eea
  ${\cal C}_{LS}$  coefficients   are  
\bea
&&{\cal C}^\rho_{10} \equiv
-x\,\left[\frac{2^{9/2}}{3^{3}}\,e_1(x)
+\frac{2^{11/2}}{3^{3/2}7^{5/2}}\,
e_2(x)
\right]\,
\nn\\
%
%
&&{\cal C}^{f_2}_{20}  \equiv  \, x^{2} \left[\frac{2^{11/2}}{3^{4}
5^{1/2}} \,e_1(x)
-\frac{2^{17/2}}{3^{3/2}5^{1/2}7^{7/2}}
\,e_2(x)
\right]
\nn \\
%
%
&&{\cal C}^{f_0}_{00}  \equiv
\frac{2^{4}}{3^{2}}
\!\left[1-\frac{2}{9}x^{2}\right]
e_1(x)
-\frac{2^{5}}{ 7^{5/2}  \sqrt{3}   }
\left[1-\frac{8}{21}x^{2}\right]
e_2(x)
\nn\\
%
%
&&{\cal C}^{a_2}_{21}  \equiv
 -\, x^{2}
\left[\frac{   2^{5}}{3^{7/2}\sqrt{5}   }
e_1(x)
-\frac{2^{7}}{7^{7/2}\sqrt{5}  }
e_2(x)
\right]
\nn\\
%
%
&&{\cal C}^{b_1}_{01} \equiv
\! -
\frac{2^{4}}{3^{5/2}}
\!\left[1-\frac{2}{9}x^{2}\right]
e_1(x)
\!+\!
\frac{2^{5}}{7^{5/2}3}
\left[1-\frac{8}{21}x^{2}\right]
e_2(x)
\nn\\
%
%
&&{\cal C}^{b_1}_{21}  \equiv
 -\, x^{2}
\left[\frac{2^{11/2}}{3^{9/2}}
e_1(x)
-\frac{2^{17/2}}{7^{7/2}3^{2}}
e_2(x)
\right]
\nn\\
%
%
&&{\cal C}^{a_1}_{01}  \equiv
\frac{2^{9/2}}{3^{5/2}}
\!\left[1-\frac{2}{9}x^{2}\right]
e_1(x)
-\frac{2^{11/2}}{7^{5/2}3}
\left[1-\frac{8}{21}x^{2}\right]
\,e_2(x)
\nn\\
%
%
&&{\cal C}^{a_1}_{21}  \equiv
 -\, x^{2}
\left[\frac{2^{5}}{3^{9/2}}
\,e_1(x)
-\frac{2^{7}5}{3^{2}7^{7/2}}
e_2(x)
\right]
\nn \\
%
%
&&{\cal C}^{h_1}_{01}  \equiv
\!
-
\frac{2^{4}}{3^{5/2}}\left[1-\frac{2}{9}x^{2}\right]
e_1(x)
\!+\!
\frac{2^{5}}{7^{5/2}3}
\left[1-\frac{8}{21}x^{2}\right]
\!e_2(x)
\nn \\
%
%
&&{\cal C}^{h_1}_{21}  \equiv
-\, x^{2}\left[\frac{2^{11/2}}{3^{9/2}}
\,e_1(x)
-\frac{2^{17/2}}{7^{7/2}3^{2}}
\,e_2(x)
\right]
\label{polinomio10}
\eea
with $x=P/\beta$; $e_1(x)=\exp\left(-x^{2}/12\right)$ and  $e_2(x)=\exp\left(-9x^{2}/28\right)$.
The decay rates are given by the following general expression
\bea
\Gamma = 2\pi^{1/2}\,\gamma^{2}
\frac{E_{B}E_{C}}{M_A}\,x\,\sum_{LS}\, {\cal C}_{LS}^{2}.
\label{taxa}
\eea
The D/S ratio for the $a_1$, $b_1$ and $h_1$ mesons is a very sensitive
experimental quantity and function of the coefficients ${\cal C}_{LS}$ in (\ref{polinomio10}).
In particular 
\bea 
{a_D\over a_S}\bigg|_{a_1\to\rho\pi} & = & \frac{ {\cal C}^{a_1}_{21} }{ {\cal C}^{a_1}_{01}       } 
\,\,\,\,\,\,\,\,;\,\,\,\,\,\,\,\, 
{a_D\over a_S}\bigg|_{b_1\to\omega\pi} = \frac{  {\cal C}^{b_1}_{21} }{  {\cal C}^{b_1}_{01}       } 
\,\,\,\,\,\,\,\,;
\,\,\,\,\,\,\,\, 
{a_D\over a_S}\bigg|_{h_1\to\rho\pi}  =  \frac{ {\cal C}^{h_1}_{21} }{  {\cal C}^{h_1}_{01} }. 
\label{d/s-eq1}
\eea 
Replacing  (\ref{polinomio10}) in (\ref{d/s-eq1}), we find
\bea 
{a_D\over a_S}\bigg|_{a_1\to\rho\pi} & =&
\frac{
 -\, x^{2}
\left\{\frac{2^{1/2}}{3^{2}} e_1(x)
-\frac{2^{5/2}3^{1/2}5}{7^{7/2}} e_2(x)
\right\} } {\left[1-\frac{2}{9}x^{2}\right] e_1(x)
-\frac{3^{3/2}2}{7^{5/2}} \left[1-\frac{8}{21}x^{2}\right] e_2(x)
}
\nn\\\nn\\\nn\\
{a_D\over a_S}\bigg|_{b_1\to\omega\pi}&=& \frac{ x^{2}
\left\{\frac{2^{3/2}}{3^{2}} e_1(x)
-\frac{2^{9/2}3^{1/2}}{7^{7/2}} e_2(x)
\right\}
}{
 \left[1-\frac{2}{9}x^{2}\right]
e_1(x)
- \frac{3^{3/2}2}{7^{5/2}}\left[1-\frac{8}{21}x^{2}\right] e_2(x)
}
\nn\\\nn\\\nn\\
{a_D\over a_S}\bigg|_{h_1\to\rho\pi} &=& \frac{ x^{2}
\left\{\frac{2^{3/2}}{3^{2}} e_1(x)
-\frac{2^{9/2}3^{1/2}}{7^{7/2}} e_2(x)
\right\} }{
 \left[1-\frac{2}{9}x^{2}\right]
e_1(x)
- \frac{3^{3/2}2}{7^{5/2}}\left[1-\frac{8}{21}x^{2}\right] e_2(x)
}.
\nn\\
\label{d/s-eq2}
\eea
In the  numerical calculation  the meson masses are 
assumed the following:
$M_{\pi}=0.139$ {\rm GeV}, $M_{\rho}=0.775$ {\rm GeV},
$M_{\omega}=0.782$ {\rm GeV}, $M_{h_1}=1.170$ {\rm GeV}, $M_{a_1}
=1.230$ {\rm GeV}, $M_{b_1}=1.229$ {\rm GeV}, $M_{f_2}=1.275$ {\rm
GeV}, $M_{f_0}=1.370$ {\rm GeV}, $M_{a_2}=1.318$ {\rm GeV}
\cite{PDG}.
To adjust the model one has to minimize $R$, defined by
\bea 
R^2=\sum_{i=1}^{7} \left[a_{i}(\gamma,\beta)-1\right]^2 
\label{hiper} 
\eea 
with  $a_{i}(\gamma,\beta)= \Gamma_{i}^{\rm thy}(\gamma,\beta) /\Gamma_{i}^{\rm exp} $.
To compare the \3p0 model with its corrected version, the minimum value for (\ref{hiper}) is obtained
for $\gamma=0.506$ and $\beta=0.397$ GeV ($R=0.559$). 
The inclusion of the   correction term
reduces the $R$ value to $0.486$ with slightly different   values for $\gamma=0.535$ and $\beta=0.387$ GeV.
A clear demonstration that the bound-state correction globally improves the fit. 

The values for $\gamma$ and $\beta$ 
are used for the seven mesons and presented  in Table  \ref{tab2} with the D/S ratios. Some comments should be made
about these results. The $b_1 \rightarrow \omega\pi$ channel is experimentally well known and in both models  $\Gamma$ and D/S ratios
are the same.
The most important discrepancy in Table \ref{tab2} is $\rho \rightarrow \pi\pi$, which is well known to be a problem relative to the
decays of P-wave $q\bar q$ mesons in the \3p0 model \cite{barnes1}.
The corrected model has an important  improvement in this channel.
In this fit the only channel where \3p0 model has better estimate  compared to the corrected model is for   $f_2$ decay.
  The $a_2$ decay, for example, is one of the three channels in a $3\pi$ mode. The total experimental decay rate 
 for the $3\pi$ mode is 75 MeV,  composed of three channels $a_2 \rightarrow \rho(770)\pi$, $a_2 \rightarrow f_2(1270)\pi$ and
$a_2 \rightarrow \rho(1450)\pi$. In our calculation,  only the first channel was evaluated and it shows a smaller contribution
to the total value of this mode in comparison to the original \3p0 model.   
The ratio for $h_1\to \rho \pi$, which
has not been measured, is theoretically close to $b_1\to \omega \pi$ to
within small phase space differences, since these are both
$^{1}P_1 \to {^{3}}S_1 +{^1}S_0$ decays. Similar to the $b_1$ decay, as expected, the $h_1$ channel 
in both models has close values for $\Gamma$ and D/S ratios. The $f_0(1370)$ decay  to $2\pi$ has a wide range
of values in PDG \cite{PDG}, consistent with both models. In our calculation $f_0(1370)$ is regarded only as 
 a $n\bar n=(u\bar u+ d\bar d)/\sqrt{2} $ state. In the literature, to the lowest order, this scalar meson has been considered
as a mixture of the scalar two gluon glueball $G$ with quarkonia states  $n\bar n $ and $s\bar s$
\cite{close1,faessler}.

\begin{table*}
\caption{Decay rates
 \3p0 \rm{(}$\gamma=0.506$ ; $\beta=0.397$ {\rm GeV )} and
$C^3P_0$  \rm{(}$\gamma=0.535$ ; $\beta=0.387$ {\rm GeV )} }
\begin{tabular}{cc|cccc|ccccccc}
&$ $ &  &$\Gamma$ (MeV)  &   &  &  & & D/S
&
\\
Decay &$$  &Exp \cite{PDG}  & \3p0 &C\3p0 &$$ & Exp \cite{PDG} &&\3p0 & & C\3p0 \\
  \hline
$\rho \rightarrow \pi\pi$& & 149.4  &81 & 111 & & && && \\
$f_2 \rightarrow \pi\pi$& & 156.7  &170 & 181 & &&& && \\
$a_2 \rightarrow \rho\pi$& &  75 ($3\pi$ mode)  &52 & 47 & &&& && \\
$a_1 \rightarrow \rho\pi$& & 250 to 600&  543   & 536& & $-0.108(16)$ &&$-0.149$ & &$-0.121$ \\
$b_1 \rightarrow \omega\pi$& & 142  &143 & 143 & &$0.277(27)$&&$0.288$ && $0.288$\\
$h_1 \rightarrow \rho\pi$& & 360  &378 & 374 & & $-$ & & $0.215$ && $0.214$\\
$f_0 \rightarrow \pi\pi$& & 126 to 460  &225 & 198 & &&& &&
\label{tab2}
\end{tabular}
\end{table*}


\section{Summary and Conclusions}

In this paper we have tested, for the first time, 
an alternative approach for meson decay, the C\3p0 model
derived from  the mapping technique, known as the Fock-Tani
formalism. The model preserves the essential predictions
of the \3p0 approach, introducing a novel feature of bound-state corrections
to the decay amplitude. The results obtained seem promising, in particular
an important improvement is seen in the decay rate of $\rho\to \pi\pi$
and in the D/S ratio of $a_1\to \rho\pi$.
From this study we should say that, although important  
as a fine-tuning to decay processes, the bound-state
correction should represent a  small contribution to dynamics due
to the extended nature of the mesons.
Future calculations can shed light to this aspect.

For  strange mesons,  glueball candidates or in the charmed sector
the inclusion of different $\beta$ values may become necessary.
 The inclusion of the full meson octet, in the evaluation of the
 bound-state kernel $\Delta$, may provide an additional adjustment in fitting  the model.
The examples studied here are encouraging, but a
more extensive survey in other
 meson sectors would be a necessary
next step.


{\bf Acknowledgements}

This research was supported by
Conselho Nacional de Desenvolvimento Cient\'{\i}fico e Tecnol\'ogico (CNPq),
Universidade Federal do Rio Grande do Sul (UFRGS) and Universidade Federal de
Pelotas (UFPel).

\end{document}